\documentclass[prb,twocolumn,groupedaddress]{revtex4}

\usepackage{graphicx}
\usepackage{dcolumn}
\usepackage{bm}

\begin{document}

\preprint{ATB-1}

\title{Spin reorientation transition in the incommensurate stripe-ordered
phase of La$_{3/2}$Sr$_{1/2}$NiO$_4$}

\author{P. G. Freeman}
\homepage{http://xray.physics.ox.ac.uk/Boothroyd}
\author{A. T. Boothroyd}
\author{D. Prabhakaran}
\affiliation{Department of Physics, Oxford University, Oxford, OX1 3PU, United
Kingdom }

\author{D. Gonz$\acute{a}$lez}
\affiliation{Instituto de Ciencia de Materials de Arag\'{o}n, CSIC-Universidad
de Zaragoza, 50009 Zaragoza, Spain }
\author{M. Enderle}\affiliation{
Institut Laue-Langevin, BP 156, 38042 Grenoble Cedex 9, France }

\date{\today}

\begin{abstract}

The spin ordering of La$_{3/2}$Sr$_{1/2}$NiO$_4$ was investigated
by magnetization measurements, and by unpolarized- and
polarized-neutron diffraction. Spin ordering with an
incommensurability $\varepsilon$ $\approx$ 0.445 is observed below
$T_{\rm SO}$ $\sim$ 80\,K. On cooling, a spin reorientation is
observed at 57\ $\pm$\ 1\,K, with the spin axes rotating from 52
$\pm$ 4$^{\circ}$ to 78 $\pm$ 3$^{\circ}$. This is the first time
a spin reorientation has been observed in a
La$_{2-x}$Sr$_x$NiO$_{4+\delta}$ compound having incommensurate
stripe order.

\end{abstract}

\pacs{}
\maketitle

Incommensurate magnetic fluctuations observed in superconducting
La$_{2-x}$Sr$_x$CuO$_4$(LSCO)\cite{super} are widely believed to
be associated with charge stripe
correlations.\cite{tranquada-Nature-1995} Assuming this to be the
case, it has been concluded that charge stripes in LSCO lie
parallel to the Cu--O bonds in the CuO planes for $x \geq$
0.053,\cite{fujita-PRB-2002} but are aligned at $45^{\circ}$ to
the Cu--O bond for $x \leq$ 0.07.
\cite{fujita-PRB-2002,superstripe} Given that the critical doping
level for superconductivivty in LSCO is $x$ =
0.053--0.056,\cite{fujita-PRB-2002} these observations suggest a
strong link between spin-charge stripe ordering and high
temperature superconductivity.

Diagonal charge stripes have also been observed in
La$_{2-x}$Sr$_x$NiO$_{4+\delta}$(LSNO),\cite{tranquada-PRL-1993}
and have been studied by
neutron\cite{neutron,yoshizawa-PRB-2000,lee-PRB-2001,kajimoto-PRB-2001,kajimoto}
and x-ray\cite{x-ray,pash-PRL-2000} diffraction for doping levels
in the range 0.135 $\leq x \leq$ 0.5. The charge stripes are found
to be ordered over a long range, with correlation lengths in
excess of 100\,\AA\ for certain levels of
doping\cite{yoshizawa-PRB-2000,pash-PRL-2000}. LSNO with $x$ = 1/3
has stripe order that is particularly stable owing to a
commensurability effect that pins the charges to the
lattice.\cite{ramirez-PRL-1996,yoshizawa-PRB-2000,kajimoto-PRB-2001}
Having charge stripes that are static and well correlated, LSNO is
a good system for studying the basic properties of spin-charge
stripes.

Recently, Lee {\it et al.}\cite{lee-PRB-2001} measured the
direction of the ordered Ni moments in the spin-ordered phase of
compounds with $x$ = 0.275 and 1/3. The angle between the spin
axes and the stripe direction was found to be larger at $x$ = 1/3
than $x$ = 0.275. Furthermore, a transition was observed at
$\sim$50\,K in the $x$ = 1/3 sample such that on cooling the spins
rotate away from the stripe direction and a further localization
of the charge ordering occurs. Lee {\it et al.} concluded from
this that an intimate connection exists between the local spin
order and the stability of the charge order.

\begin{figure}[!ht]
\begin{center}
\includegraphics{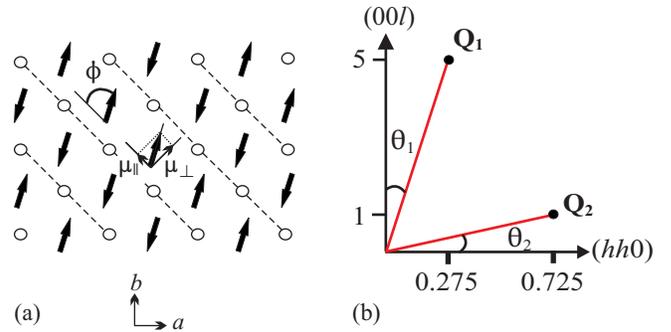}
\caption[Checkerboard and peak positions]{(a) Ideal checkerboard
2-D charge ordering in the $ab$ plane. Circles represent holes
residing on Ni sites, while the solid arrows represent the spins
of the Ni$^{2+}$ sites. Dashed lines indicate the charge domain
walls common to both checkerboard and stripe ordering. The
commensurate spin ordering shown here is not realized in practise
in La$_{3/2}$Sr$_{1/2}$NiO$_{4+\delta}$. The observed spin order
in  La$_{3/2}$Sr$_{1/2}$NiO$_{4+\delta}$ is similar to that shown,
but is actually incommensurate in the direction perpendicular to
the stripes. The components of the ordered moment parallel
($\mu$$_\|$) and perpendicular ($\mu$$_\bot$) to the stripe
direction are shown, and $\phi$ denotes the angle between the spin
axis and the stripe direction. (b) Diagram of the ($h,\ h,\ l$)
plane in reciprocal space. {\bf Q$_1$} and {\bf Q$_2$} are the
scattering vectors of the two magnetic peaks investigated in this
work. } \label{fig:fig1}
\end{center}
\end{figure}

Early work on La$_{3/2}$Sr$_{1/2}$NiO$_{4+\delta}$ by electron
diffraction\cite{chen-PRL-1993} suggested that the case of
half-doping is another very stable charge-ordering phase. This
charge ordering has a 2-D checkerboard-like pattern as shown in
figure 1(a). Recently, Kajimoto {\it et al.}\cite{kajimoto}
studied the spin and charge order in a single crystal of
La$_{3/2}$Sr$_{1/2}$NiO$_4$ by neutron diffraction. They observed
checkerboard ordering below $\sim$480\,K, but they also found an
incommensurate charge-stripe ordering phase below $T_{\rm
ICO}$$\sim$ 180\,K coexisting with the checkerboard ordering. Spin
ordering was reported to be found only in association with the
incommensurate component of the charge ordering.

Here we describe a study of spin-ordering in
La$_{3/2}$Sr$_{1/2}$NiO$_{4+\delta}$ employing magnetometry and
polarized- and unpolarized-neutron diffraction. Our results are
consistent with those of Kajimoto {\it et al.},\cite{kajimoto} but
in addition we observed a spin reorientation transition similar
to, but more pronounced than that observed in
La$_{5/3}$Sr$_{1/3}$NiO$_4$. This is the first time a spin
reorientation associated with an incommensurate spin ordering has
been observed in a LSNO material. The spin reorientation is
signalled by a sharp anomaly in the magnetization at $\sim$57\,K,
and an analysis of polarized-neutron diffraction data shows that
on cooling through 57\,K the Ni spins rotate away from the stripe
direction.

\begin{figure}[!t]
\begin{center}
\includegraphics{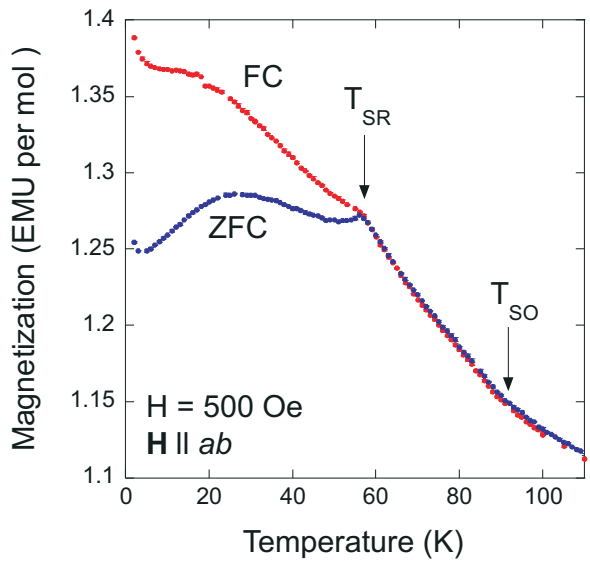}
\caption[SQUID]{FC and ZFC magnetization data for
La$_{3/2}$Sr$_{1/2}$NiO$_{4+\delta}$. Arrows indicate the
spin-ordering temperature, $T_{\rm SO}$, and the
spin-reorientation temperature, $T_{\rm SR}$. } \label{fig:SQUID}
\end{center}
\end{figure}

Two single crystals of La$_{3/2}$Sr$_{1/2}$NiO$_{4+\delta}$ were
used for these experiments. The crystal used for the magnetization
measurements had dimensions $\sim$$7\times5\times2$\,mm$^{3}$ and
that for neutron diffraction was a rod of 7--8\,mm diameter and
$\sim$35\,mm in length. Both these samples were grown from the
same starting powder by the floating-zone technique.\cite{Prab}
The oxygen excess was determined by thermogravimetric analysis to
be $\delta$ = 0.02  $\pm$ 0.01.

Magnetization data were collected with a SQUID magnetometer
(Quantum Design) with the field applied parallel to the $ab$ plane
of the crystal. We carried out DC measurements by cooling in an
applied field of 500\,Oe (FC) and also by cooling in zero field
and measuring while warming in a field of 500\,Oe (ZFC). Figure 2
shows the temperature variation of the FC and ZFC magnetization. A
small change in slope at 90 $\pm$ 5\,K marks the spin ordering
temperature, in agreement with previous
observations.\cite{yoshizawa-PRB-2000}
 A clearer anomaly is seen at $T_{\rm SR}$ = 57\,K,
where the FC and ZFC curves sharply separate. Below 5\,K both the
FC and ZFC curves begin to rise for reasons that are not yet
understood, but could be due to a small amount of paramagnetic
impurity in the crystal. We measured the FC magnetization parallel
to the crystal $c$ direction (not shown) and found no anomaly
around $T_{\rm SR}$, thus indicating that the transition involves
the in-plane components of the spins.

In what follows we describe neutron diffraction studies aimed at
identifying the nature of the transition at $T_{\rm SR}$. Referred
to the tetragonal unit cell of LSNO, with unit cell parameters $a$
$\approx$ 3.8\,\AA, $c$ $\approx$ 12.7\,\AA, charge ordering peaks
due to structural distortions are located at ($h{\pm\varepsilon}$,
$k{\pm\varepsilon}$, $l$) positions in reciprocal space. The spin
order peaks are located at ($h$+1/2${\pm\varepsilon}$/2,\
$k$+1/2${\pm\varepsilon}$/2,\ $l$), the strongest being when $l$
is odd.\cite{lee-PRB-2001} To a first approximation the
relationship between the incommensurability, $\varepsilon$, and
the hole concentration, $n_h = x$ + 2 $\delta$, is simply
$\varepsilon$ = $n_h$. In general, however, competition between
apparently different wavevectors for the spin and charge
ordering\cite{kajimoto-PRB-2001} and the separate issue of oxygen
non-stoichiometry\cite{jest-PRB-1999,wochner-PRB-1998,chen-PRL-1993}
causes small deviations from the $\varepsilon = n_h$ rule.

The neutron experiments were performed on the triple-axis
spectrometer IN20 at the Institut Laue-Langevin. The energies of
the incident and elastically-scattered neutrons were selected by
Bragg reflection from an array of either pyrolytic graphite (PG)
or Heusler alloy crystals, for unpolarized- and polarized-neutron
scattering respectively. A PG filter was present after the sample
and before the analyzer to suppress higher-order harmonic
scattering.  Most of the data were obtained with initial and final
neutron wavevectors of 2.66\,\AA$^{-1}$, with some further data
taken with a neutron wavevector of 3.54\,\AA$^{-1}$. We mounted
the sample with the [001] and [110] crystal directions in the
horizontal scattering plane, and performed scans in reciprocal
space either parallel to the ($h$,\ $h$,\ 0) direction at constant
$l$, or parallel to the (0,\ 0,\ $l$) direction at constant $h$.

We began by performing polarized-neutron diffraction measurements with the neutron
polarization {\bf P} parallel to the scattering vector {\bf Q}, maintained by adjusting currents
in a Helmholtz coil mounted around the sample. In this configuration a neutron's spin is
flipped during a magnetic scattering process, but remains unchanged when scattered by a
non-magnetic process, e.g. from a distortion of the lattice. Thus, by measuring the spin-
flip (SF) and non-spin-flip (NSF) channels one can identify whether the origin of the
scattering is magnetic or not.

We observed charge-order diffraction in the vicinity of the (1.5,
1.5, $l$) line in reciprocal space. At $T$ = 10\,K the peak shape
observed in ($h$,\ $h$,\ 0) scans was consistent with that
observed by Kajimoto {\it et al.}, with a central peak at $h$ =
1.5 and two satellites at $h \approx$ 1.45 and 1.55. We observed
no variation of this scattering with $l$ over a single Brillouin
zone, showing that both the commensurate and incommensurate
components of charge ordering are completely two-dimensional.

\begin{figure}[!ht]
\begin{center}
\includegraphics{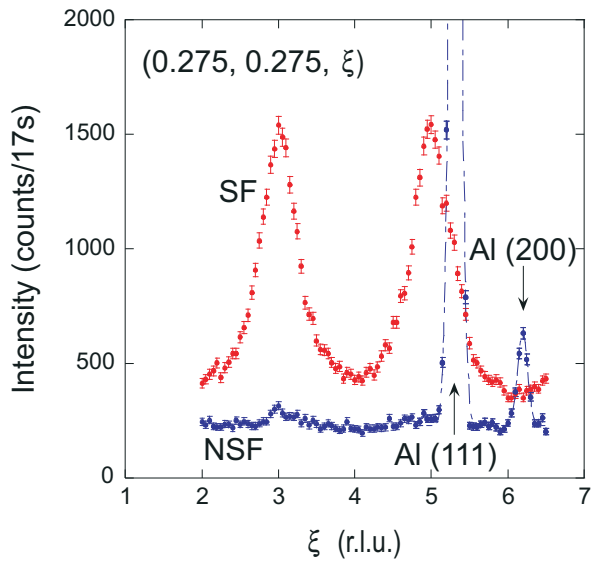}
\caption[lscan]{Spin-flip (SF) and non-spin-flip (NSF) diffraction
from La$_{3/2}$Sr$_{1/2}$NiO$_4$ at $T$ = 10 K. The scan is
parallel to (0,\ 0,\ $l$) and passes through the magnetic order
peak (0.275,\ 0.275,\ 1). No correction has been made for the
imperfect polarization of the neutron beam. The sharp peaks in the
NSF channel at $l \approx$ 5.3 and 6.2 are due to diffraction from
the Al sample mount.} \label{lscan}
\end{center}
\end{figure}

We observed spin-order diffraction peaks at
($h$+1/2$\pm\varepsilon$/2, $h$+1/2$\pm\varepsilon$/2,\ $l$)
positions with $l$ odd and $\varepsilon$ = 0.445 $\pm$ 0.005.
Figure 3 shows a scan parallel to (0,\ 0,\ $l$). The widths of the
peaks in this scan convert to a correlation length along the $c$
axis of 16.4 $\pm$ 0.3\,\AA. For comparison, the in-plane
correlation length in the direction perpendicular to the stripes
is 78.3 $\pm$ 1.3\,\AA. There were no peaks at $l$ = even
positions, as can be seen in figure 3. There were also no
spin-order peaks at ($h\pm$1/4, $h\pm$1/4,\ l) positions,
confirming the lack of any commensurate Fourier component in the
spin ordering, in agreement with the observations of Kajimoto {\it
et al.}.\cite{kajimoto}

\begin{figure}[!ht]
\begin{center}
\includegraphics{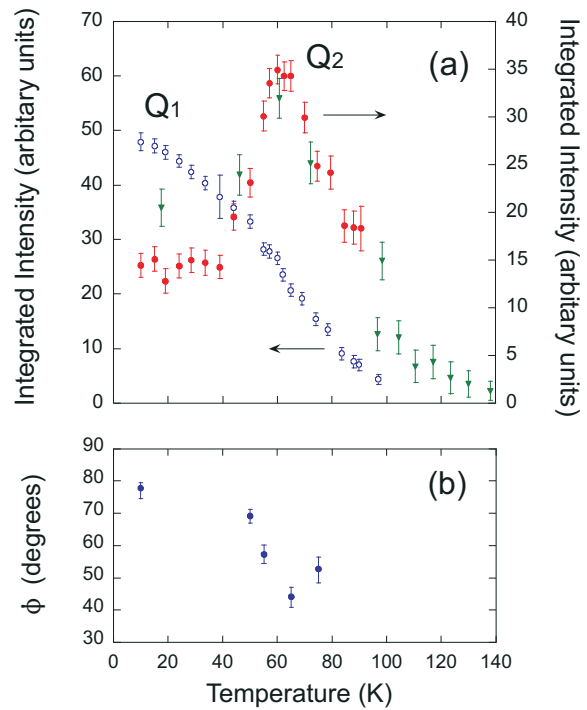}
\caption[peak intensity and spin direction]{(a) The peak
intensities of the spin peaks ${\bf Q}_1=(0.275,0.275,5)$ and
${\bf Q}_2=(0.725,0.725,1)$
--- see figure 1(b). Circle symbols represent data taken with a
neutron wavevector of 3.54\,\AA$^{-1}$ and triangles represent
2.66\,\AA$^{-1}$. Filled symbols are data taken at {\bf Q$_2$} and
unfilled symbols at {\bf {Q$_1$}}, with the arrows indicating the
relevant scales for each {\bf Q}. (b) The temperature dependence
of the angle $\phi$ between the spin axis and the stripe direction
obtained from polarized-neutron analysis. } \label{fig:polar}
\end{center}
\end{figure}

Figure 4a show the temperature dependence of the integrated
intensity of two magnetic reflections, {\bf Q$_1$} =
(0.275,0.275,5) and {\bf Q$_2$} = (0725,0.725,1), measured in
scans parallel to $(h, h, 0)$. The spin ordering transition is
seen to be rather sluggish, but a sharpening of the peaks below
$T_{\rm SO}$ $\simeq$ 80\,K (not shown) suggests that this is
where long-range ordering extending over many unit cells sets in.
However short-range correlations persist above 100\,K. On cooling
below $T_{\rm SO}$, the peak at {\bf Q$_1$} increases
monotonically, while that at {\bf Q$_2$} first increases, then
reaches a maximum at $\sim$60\,K, then decreases again. This
anomalous behaviour correlates with the transition observed in the
magnetization, Fig. 2. To understand this behaviour we must recall
that magnetic neutron diffraction is sensitive to spin components
perpendicular to {\bf Q}. As shown in figure 1(b), {\bf Q$_1$} is
aproximately parallel to $(0, 0, l)$ while {\bf Q$_2$} is
approximately parallel to $(h,h,0)$. Therefore, to a good
approximation, the scattering at {\bf Q$_1$} is from the total
in-plane spin moment, irrespective of its direction, while that at
{\bf Q$_2$} comes mainly from the spin components parallel to the
stripes and along the $c$ axis. The data in fig. 4a therefore
imply that below $\sim$57\,K the spins rotate away from the stripe
direction.

\begin{table*}
\begin{ruledtabular}
\begin{tabular}{ccccc}
&{\bf P} configuration & &{\bf {Q$_1$}}& {\bf {Q$_2$}}\\
\hline
\\
1)&{\bf P} $\|$ {\bf Q}& SF & $\mu$$_\bot^{2}$ $\cos$$^{2}$$\theta$$_1$ + $\mu$$_\|^{2}$ + $\mu$$_C^{2}$ $\sin$$^{2}$$\theta$$_1$& $\mu$$_\bot^{2}$ $\sin$$^{2}$$\theta$$_2$ + $\mu$$_\|^{2}$ + $\mu$$_C^{2}$ $\cos$$^{2}$$\theta$$_2$\\
&& NSF & NM & NM \\
2)&{\bf P} $\bot$ {\bf Q} ({\bf P} in horiz. plane)& SF & $\mu$$_\|^{2}$ &$\mu$$_\|^{2}$ \\
&& NSF & $\mu$$_\bot^{2}$ $\cos$$^{2}$$\theta$$_1$+ $\mu$$_C^{2}$ $\sin$$^{2}$$\theta$$_1$ + NM & $\mu$$_\bot^{2}$ $\sin$$^{2}$$\theta$$_2$ + $\mu$$_C^{2}$ $\cos$$^{2}$$\theta$$_2$ + NM\\
3)&{\bf P} $\bot$ {\bf Q} ({\bf P} vertical)& SF & $\mu$$_\bot^{2}$ $\cos$$^{2}$$\theta$$_1$+ $\mu$$_C^{2}$ $\sin$$^{2}$$\theta$$_1$ & $\mu$$_\bot^{2}$ $\sin$$^{2}$$\theta$$_2$ + $\mu$$_C^{2}$ $\cos$$^{2}$$\theta$$_1$\\
&& NSF &$\mu$$_\|^{2}$ + NM & $\mu$$_\|^{2}$ + NM \\
\end{tabular}
\end{ruledtabular}
\caption{\label{tab:table1} Expressions for the intensity of SF
and NSF scattering for different orientations of the neutron
polarization {\bf P} relative to the scattering vector {\bf Q} (NM
is the non-magnetic scattering). }
\end{table*}

To fully analyze the direction of the spins over this temperature
range we varied the direction of the neutron polarization {\bf P}
relative to the scattering vector {\bf Q}. Three configuration
were used: 1) {\bf P} $\parallel$ {\bf Q}, 2) {\bf P} $\bot$ {\bf
Q} but in the scattering plane, and 3) {\bf P} $\bot$ {\bf Q} but
out of the scattering plane. (Recall that [110] and [001] are the
crystal directions in the scattering plane.) As mentioned before,
neutrons scatter via magnetic interactions from spin components
perpendicular to {\bf Q}, and SF scattering is due to spin
components perpendicular to {\bf P}. Therefore, analysis of
configurations 1), 2) and 3) leads to the direction of the ordered
moment.\footnote{Actually, 2 out of the 3 configurations are
sufficient.} Table 1 summarizes the relations between the observed
intensities at {\bf Q$_1$} and {\bf Q$_2$} and the ordered spin
components parallel ($\mu$$_\|$) and perpendicular ($\mu$$_\bot$)
to the stripe direction (see figure 1(a)), and parallel to the $c$
axis ($\mu_c$) .

We corrected the data to take into account the different
background count rates in the SF and the NSF channels, and to
correct for the imperfect spin polarization of neutron beam. The
latter was calculated from the flipping ratio 18.5 $\pm$ 1.8
measured on a magnetic Bragg peak.

A full analysis of the intensities at the {\bf Q$_1$} and {\bf
Q$_2$} points determined the spin orientations at $T$ = 10\,K and
$T$ = 65\,K. These analyses showed that the non-magnetic (NM) and
$\mu_c$ count rates were zero, to within error for both
temperatures. Thus, it is reasonable to assume that the spins lie
within the $ab$ plane over the whole temperature range, consistent
with the lack of any anomaly observed at $T_{\rm SR}$ in the
magnetization data measured with ${\bf H}\ \|\  c$. At the other
temperatures we deduced the spin orientation within the $ab$ plane
from one {\bf Q} position ($T$= 55\,K from {\bf Q$_1$}, $T$ = 50
and 75\,K from {\bf Q$_2$}.) From these data we find that the
spins reorientate from an angle of 52 $\pm$ 4$^{\circ}$ to the
stripe direction at $T$ = 75\,K, to 78 $\pm$ 3$^{\circ}$ at $T$ =
10\,K, as shown in figure 4(b). The transition is seen to occur
quite rapidly between 50 and 65\,K, although the reorientation is
not complete at 50\,K.

How does the spin reorientation described here in the $x$ = 1/2
crystal compare with that reported by Lee {\it et
al.}\cite{lee-PRB-2001} for $x$ = 1/3? In both compositions the
spin reorientations occur at similar temperatures ($\sim$50\,K for
$x$ = 1/3, $\sim$57\,K for $x$ = 1/2) and the spins rotate in the
same sense, namely away from the stripe direction, on cooling. On
the other hand, the angle through which the spins rotate is
greater for $x$ = 1/2 ($\sim$26$^{\circ}$) than for $x$ = 1/3
($\sim$13$^{\circ}$), and the final low temperature spin direction
in $x$ = 1/2 is almost perpendicular to the stripes ($\phi$ =
78$^{\circ}$) compared with 53$^{\circ}$ for $x$ = 1/3.
Considering also the spin direction $\phi$ = 27$^{\circ}$ observed
for $x$ = 0.275,\cite{lee-PRB-2001} for which there is no
reorientation transition, we find that as the doping level
increases the angle from the stripe direction increases. However,
the trend in the spin orientation also correlates with the
increasing stability of the charge ordering as measured by the
charge-ordering
temperatures\cite{yoshizawa-PRB-2000,lee-PRB-2001,kajimoto}
($T_{\rm CO}$($x$ = 0.275) $\sim 200$\,K, $T_{\rm CO}$($x$ = 1/3)
$\sim 240$\,K, $T_{\rm CO}$($x$ = 1/2) $\sim 480$\,K). Therefore,
measurements on more compositions of
La$_{2-x}$Sr$_x$NiO$_{4+\delta}$ are needed to establish whether
it is the doping level or the stability of the charge order that
is more important in determining the spin orientation.

The occurrence of spin reorientation transitions in the $x$ = 1/3
and $x$ = 1/2 but not in $x$ = 0.275 might suggest that these
transitions are linked to the stability of the charge order.
Irrespective of this, what the present study has shown is that the
occurrence of a spin reorientation transition does not depend on
the stripe order being commensurate with lattice. Indeed, the spin
reorientation transition is even more striking in the $x$ = 1/2
composition, for which the spin order is incommensurate, than in
$x$ = 1/3, for which it is commensurate.

These observations emphasize the importance of couplings between
spin and charge degrees of freedom in stripe-ordered systems, and
suggest that this coupling is particularly strong in
La$_{3/2}$Sr$_{1/2}$NiO$_4$.

This work was supported in part by the Engineering and Physical
Sciences Research Council of Great Britain.


\begin{references}

\bibitem{super}
T. R. Thurston {\it et al.}, Phys. Rev. B {\bf 40}, 4585 (1989);
S-W. Cheong {\it et al.}, Phys. Rev. Lett. {\bf 67}, 1791 (1991);
T. E. Mason, G. Aeppli, and H. A. Mook, Phys. Rev. Lett. {\bf 68},
1414 (1992); T. R. Thurston {\it et al.}, Phys. Rev. B {\bf 46},
9128 (1992);

\bibitem{tranquada-Nature-1995}
J. M. Tranquada {\it et al.}, Nature (London) {\bf 375}, 561
(1995).

\bibitem{fujita-PRB-2002}
M. Fujita {\it et al.}, Phys. Rev. B {\bf 65}, 064505 (2002).

\bibitem{superstripe}
S. Wakimoto {\it et al.}, Phys. Rev. B {\bf 60}, R769 (1999); S.
Wakimoto {\it et al.}, Phys. Rev. B {\bf 61}, 3699 (2000); M.
Matsuda {\it et al.}, Phys. Rev. B {\bf 62}, 9148 (2000).


\bibitem{tranquada-PRL-1993}
J. M. Tranquada, D. J. Buttrey, D. E. Rice, Phys. Rev. Lett. {\bf
70}, 445 (1993)

\bibitem{neutron}
S. M. Hayden {\it et al.}, Phys. Rev. Lett. {\bf 68}, 1061 (1992);
V. Sachan {\it et al.}, Phys. Rev. B {\bf 51}, 12742 (1995); J. M.
Tranquada, D. J. Buttrey, and V. Sachan, Phys. Rev. B {\bf 54},
12318 (1996);

\bibitem{yoshizawa-PRB-2000}
H. Yoshizawa {\it et al.}, Phys. Rev. B {\bf 61}, R854 (2000)


\bibitem{lee-PRB-2001}
S-H. Lee {\it et al.}, Phys. Rev. B {\bf 63}, 060405 (2001)


\bibitem{kajimoto-PRB-2001}
R. Kajimoto {\it et al.}, Phys. Rev. B {\bf 64}, 144432 (2001)

\bibitem{kajimoto}
R. Kajimoto {\it et al.}, cond-mat/020228 (2002)

\bibitem{x-ray}
E. D. Isaacs {\it et al.}, Phys. Rev. Lett. {\bf 72}, 3421 (1994);
A. Vigliante {\it et al.}, Phys. Rev. B {\bf 56}, 8248 (1997)

\bibitem{pash-PRL-2000}
Yu. G. Pashkevich {\it et al.}, Phys. Rev. Lett. {\bf 84}, 3919
(2000)

\bibitem{ramirez-PRL-1996}
A. P. Ramirez {\it et al.}, Phys. Rev. Lett. {\bf 76}, 447 (1996)


\bibitem{chen-PRL-1993}
C. H. Chen, S-W. Cheong, and A. S. Cooper, Phys. Rev. Lett. {\bf
71}, 2461 (1993)

\bibitem{Prab}
D. Prabhakaran, P. Isla, and A. T. Boothroyd, J. Cryst. Growth
{\bf 237}, 815 (2002)


\bibitem{jest-PRB-1999}
Th. Jest\"{a}dt {\it et al.}, Phys. Rev. B {\bf 59}, 3775 (1999)

\bibitem{wochner-PRB-1998}
P. Wochner {\it et al.}, Phys. Rev. B {\bf 57}, 1066 (1998)



\end{references}
\end{document}